\documentclass[sigconf]{acmart}
\usepackage{diagbox}
\usepackage{graphicx} 
\usepackage{mathbbol}
\usepackage{amsmath}
\usepackage{multirow}
\usepackage{subfig}
\usepackage{colortbl}

\AtBeginDocument{%
  }
\copyrightyear{2024}
\acmYear{2024}
\setcopyright{acmlicensed}

\acmConference[KDD '24]{Proceedings of the 30th ACM SIGKDD Conference on Knowledge Discovery and Data Mining}{August 25--29, 2024}{Barcelona, Spain}
\acmBooktitle{Proceedings of the 30th ACM SIGKDD Conference on Knowledge Discovery and Data Mining (KDD '24), August 25--29, 2024, Barcelona, Spain}
\acmISBN{979-8-4007-0490-1/24/08}
\acmDOI{10.1145/3637528.3671655}

\begin{document}

\title{From Variability to Stability: Advancing RecSys Benchmarking Practices}

\author{Valeriy Shevchenko} 
\affiliation{
 \institution{Skoltech}
 \city{Moscow}
 \country{Russian Federation}
}
\email{valeriy.shevchenko@skoltech.ru}

\author{Nikita Belousov}
\affiliation{
 \institution{Skoltech}
 \city{Moscow}
 \country{Russian Federation}
}
\email{nikita@nokiroki.ru}

\author{Alexey Vasilev}
\affiliation{
  \institution{Sber AI Lab}
  \city{Moscow}
  \country{Russian Federation}
}
\email{alexxl.vasilev@yandex.ru}

\author{Vladimir Zholobov}
\affiliation{
 \institution{Skoltech}
 \city{Moscow}
 \country{Russian Federation}
}
\affiliation{
 \institution{MIPT}
 \city{Moscow}
 \country{Russian Federation}
}
\email{v.zholobov@skoltech.ru}

\author{Artyom Sosedka}
\affiliation{
  \institution{Sber AI Lab}
  \city{Moscow}
  \country{Russian Federation}
}
\email{m1801239@edu.misis.ru}

\author{Natalia Semenova}
\affiliation{
  \institution{AIRI}
  \city{Moscow}
  \country{Russian Federation}
}
\affiliation{
  \institution{Sber AI Lab}
  \city{Moscow}
  \country{Russian Federation}
}
\email{semenova.bnl@gmail.com}

\author{Anna Volodkevich}
\affiliation{
  \institution{Sber AI Lab}
  \city{Moscow}
  \country{Russian Federation}
}
\email{volodkanna@yandex.ru}

\author{Andrey Savchenko}
\affiliation{
  \institution{Sber AI Lab}
  \city{Moscow}
  \country{Russian Federation}
}
\affiliation{
 \institution{HSE University}
 \city{Nizhny Novgorod}
 \country{Russia}
}
\email{avsavchenko@hse.ru}

\author{Alexey Zaytsev}
\affiliation{
 \institution{Skoltech}
 \city{Moscow}
 \country{Russian Federation}
}
\affiliation{
 \institution{BIMSA}
 \city{Beijing}
 \country{China}
}
\email{likzet@gmail.com}


\renewcommand{\shortauthors}{Valeriy Shevchenko, et al.}


\begin{abstract}
In the rapidly evolving domain of Recommender Systems (RecSys), new algorithms frequently claim state-of-the-art performance based on evaluations over a limited set of arbitrarily selected datasets. However, this approach may fail to holistically reflect their effectiveness due to the significant impact of dataset characteristics on algorithm performance. Addressing this deficiency, this paper introduces a novel benchmarking methodology to facilitate a fair and robust comparison of RecSys algorithms, thereby advancing evaluation practices. By utilizing a diverse set of $30$ open datasets, including two introduced in this work, and evaluating $11$ collaborative filtering algorithms across $9$ metrics, we critically examine the influence of dataset characteristics on algorithm performance. We further investigate the feasibility of aggregating outcomes from multiple datasets into a unified ranking. Through rigorous experimental analysis, we validate the reliability of our methodology under the variability of datasets, offering a benchmarking strategy that balances quality and computational demands. This methodology enables 
a fair yet effective means of evaluating RecSys algorithms, providing valuable guidance for future research endeavors.

\end{abstract}



\maketitle

\section{Introduction}

Recommender systems have become the backbone of personalizing user experiences across diverse online platforms. 
By suggesting movies, recommending products, and curating news feeds~\cite{krichene2020sampled}, RecSys is a key machine learning technology widely used in many applications.
Their impact drives ongoing development in both academia and industry, resulting in the introduction of numerous RecSys algorithms each year~\cite{sun2023take}.

With this ongoing expansion, there is a growing need for tools that enable reproducible evaluation, allowing researchers to assess new methods alongside well-established baselines~\cite{sun2020we, ferrari2021troubling, hidasi2023effect}. 
While several frameworks \cite{anelli2021elliot, zhao2022recbole, sun2022daisyrec}
excel in conducting a rigorous evaluation of RecSys algorithms on a specific dataset, selecting the best-performing models across multiple problems remains challenging.
Results vary significantly based on the considered dataset, and what works well in one context may perform poorly in another~\cite{chin2022datasets}.
This variability often results in inconsistent conclusions from evaluation studies, highlighting the importance of comparing algorithms across datasets with various data characteristics. 
On the other hand, extensive evaluation with dozens of datasets uses large amounts of computational resources --- and harms both the environment and opportunities for small research labs. 
With researchers seeking universally effective algorithms across different recommendation tasks, businesses question algorithms' performance on datasets that reflect their specific industry domain or characteristics, trying to shorten time-to-production for RecSys. 

However, in contrast with other machine learning subdomains like time series classification~\cite{bellogin2017statistical} and NLP~\cite{rofin2022vote}, the field of RecSys lacks an accepted performance aggregation method across multiple datasets.
Furthermore, there is limited research dedicated to comparing and contrasting different recommendation datasets, understanding their impact on the performance of RecSys algorithms, and identifying datasets with similar characteristics. 

To deal with these problems, we develop a comprehensive benchmark methodology that can reliably rank RecSys methods based on their performance across various problems using offline evaluation, overcoming the limitations of current practices.
Our approach confidently determines whether a specific top-1 model can excel universally or within particular domains defined by dataset characteristics. 
While providing reliable results, we use only a small number of datasets, enabling robust and efficient comparison.

Our contributions include:
\begin{itemize}
    \item A benchmarking methodology tailored to the RecSys domain with a clear evaluation protocol and hyperparameter tuning~\footnote{To guarantee the reproducibility of our results, all code and datasets employed in experiments are available in the GitHub repository \url{https://github.com/nokiroki/recsys-evaluation-of-dozens-datasets}.}.
    
    \item Utilization of 30 public datasets for benchmarking. Among them, there are two new large-scale open-source datasets from distinct RecSys domains (music and e-commerce)  ~\footnote{\url{https://www.kaggle.com/datasets/alexxl/zvuk-dataset} and \url{https://www.kaggle.com/datasets/alexxl/megamarket}}.
    
    \item Comparative analysis of various metrics aggregation methods and their robustness stress tests to identify the most suitable approach for RecSys multi-dataset benchmarking.
    
    \item Investigation into the relationship between specific dataset characteristics and recommendation quality and identification of dataset clusters with similar characteristics.

    \item Efficient comparison procedure that uses only 6 datasets but provides similar ranking due to reasonable selection of benchmarking datasets based on clustering. 
    
    \item Identification of the top-performing algorithms from a pool of 11 frequently used approaches based on principled metrics aggregation across multiple scenarios.

\end{itemize}



\section{Related work}

\textbf{\textit{RecSys evaluation.}} Recommender systems continue to be a dynamic research area.
Traditional techniques like Neighborhood-based models \cite{sarwar2001item} and Matrix Factorization~\cite{5197422} remain reliable baselines.
However, incorporating Deep Neural Networks has notably advanced RecSys, significantly enriching the domain~\cite{zhang2019deep}. 
This variety leads to the development of open-source libraries and tools to address diverse application needs. 
Among the noteworthy ones,  DeepRec~\cite{zhang2019deeprec}, Implicit~\cite{Implicit}, LightFM~\cite{Kula15}, NeuRec~\cite{NeuRec}, RecBole~\cite{zhao2021recbole}, RecPack~\cite{RecPack2022} and Replay~\cite{replay-rec} offer realizations of popular recommendation algorithms. 

Offline evaluation remains essential in RecSys research as it provides a reliable and cost-effective approach to assess algorithm performance. 
It is particularly suitable for researchers who are developing new models.
As a part of offline evaluation, the variety in the field leads to the need for rigorous and reproducible evaluation methodologies.
Notable studies such as Elliot~\cite{anelli2021elliot}, Recbole~\cite{zhao2022recbole}, and DaisyRec~\cite{sun2022daisyrec} introduced comprehensive evaluation frameworks in both reproducing and benchmarking recommendation models. These frameworks offer a rich array of options for pre-filtering, data splitting, evaluation metrics, and hyperparameter tuning across a broad spectrum of popular recommender models.  
Notably, Elliot uniquely provides statistical tests for robustly analyzing the final results, adding a layer to the evaluation process.

\textbf{\textit{RecSys datasets.}} Dozens of public datasets from diverse domains are available for constructing and evaluating recommender systems. The research~\cite{zhao2022revisiting} shows that most studies utilize, on average, $2.9$ datasets, with dataset selection and preprocessing affecting evaluation outcomes significantly.
Different data filtering techniques can change data characteristics, leading to varied performance rankings.
Deldjoo et al.~\cite{deldjoo2020dataset, deldjoo2021explaining}
investigated how data properties impact recommendation accuracy, fairness, and vulnerability to shilling attacks, highlighting the importance of data understanding in enhancing system performance.
The paper~\cite{chin2022datasets} emphasized the crucial role of dataset diversity in RecSys algorithm evaluations, showing that dataset choice significantly affects evaluation conclusions. These findings collectively underscore the need for considering dataset variability in future research to enhance the reliability of recommender system evaluations.

\textbf{\textit{Aggregating methods.}} When introducing a novel machine learning approach, it is important to rigorously compare its performance against existing methods across a comprehensive set of relevant tasks to determine its standing relative to the current state-of-the-art.  
However, drawing conclusions about the superior algorithm based on the outcomes from multi-dataset benchmarks can be challenging.

Various techniques have been developed to yield concise summaries to address this challenge.
One basic method involves mean aggregation, assuming uniformity across task metrics~\cite{colombo2022infolm}. However, this can lead to biases when metrics vary significantly~\cite{niessl2022over}. 
The Dolan-Moré performance profiles, initially developed for optimization algorithm benchmarks~\cite{dolan2002benchmarking}, have gained traction for evaluating machine learning algorithm efficacy across diverse problems~\cite{belyaev2016gtapprox}.
Unlike mean aggregation, Dolan-Moré curves consider the distribution of performance values, offering insight into how frequently and significantly an algorithm excels. 
Similarly, the Critical Difference (CD) diagram~\cite{demvsar2006statistical} is frequently used to compare algorithms across multiple tasks.
This method of presenting results has become broadly accepted~\cite{middlehurst2021hive},
providing both groupwise and pairwise comparisons. 
Groupwise comparisons are achieved by ordering all methods based on the mean rank of relative performance on each task. 
Pairwise comparisons are based on a critical difference in this mean rank or, later~\cite{benavoli2016should}, the Wilcoxon signed-rank test~\cite{wilcoxon1992individual} with multiple test corrections. 

VOTE'N'RANK~\cite{rofin2022vote} is another framework proposed for ranking systems in multitask benchmarks rooted in the principles of social choice theory.
The framework employs scoring and majority-relation-based rules, such as Plurality, Dowdall, Borda, Copeland, and Minimax, to ensure a more comprehensive evaluation.

\textbf{\textit{Benchmarking}} is a fundamental practice in machine learning, crucial for measuring progress through datasets, metrics, and aggregation methodologies to evaluate system performance. These benchmarks are crucial for comparing new algorithms with established ones to identify the most effective models for practical use.

Performance benchmarks are essential across various fields. For example, 
the ILSVRC (ImageNet Large Scale Visual Recognition Challenge)~\cite{russakovsky2015imagenet} consider object classification and detection with extensive image datasets and unique metrics for each task. 
In natural language processing (NLP), GLUE~\cite{wang2018glue} and its derivatives~\cite{wang2019superglue} benchmark models across diverse tasks, ranking them based on mean score values.
One example in the AutoML domain is AMLB~\cite{gijsbers2022amlb}, which emphasizes multitask evaluation via mean ranking.

The research~\cite{anelli2022top} offers an in-depth and reproducible evaluation of ten collaborative filtering algorithms, employing a Borda count ranking method to aggregate accuracy results from various datasets and metrics. The study emphasizes that, although this method provides valuable insights, it necessitates careful interpretation due to biases favouring algorithms that perform well in correlated metrics.

\begin{figure*}[!th]
    \centering
    \includegraphics[width=0.95\textwidth]{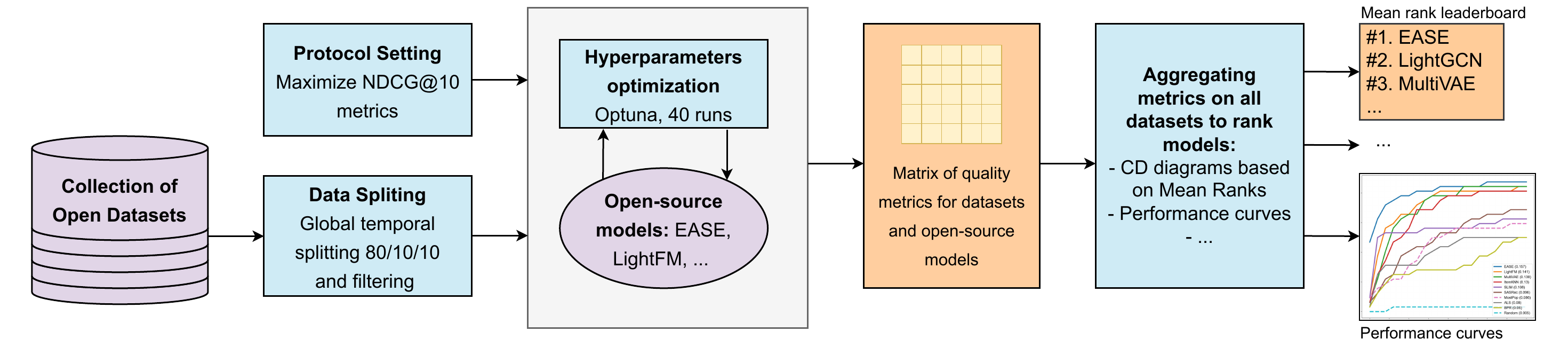}
    \caption{Benchmarking methodology for ranking algorithms. Our main innovations are the curated list of datasets that enable the option of comparison of pairs of models and aggregation strategies that provide principled ranking of approaches w.r.t. various criteria.}
    \label{fig:pipeline_scheme}
\end{figure*}

To the best of our knowledge, BARS~\cite{zhu2022bars} is the most advanced benchmarking initiative focused on RecSys. 
Although BARS establishes an open benchmark with standardized evaluation protocols, it presents certain limitations.
For instance, it restricts itself to only three datasets dedicated to the singular challenge of top-N recommendation, maintaining discrete leaderboards for each dataset.
Such an approach, devoid of a multi-dataset scoring mechanism, inhibits discerning truly adaptive and universal models — covering this gap might offer significant insights for researchers.

\section{Methodology}

Our aim is to present a robust and efficient benchmarking methodology tailored for the RecSys domain. 
We align our experimental setup with online evaluation, replicating real-time recommendation scenarios while 
ensuring the reproducibility of our benchmarking results. 

To achieve our goal, we collect a diverse set of open-source datasets and establish
a robust pipeline that incorporates predefined procedural steps. 
Additionally, we integrate 11 RecSys algorithms from various open-source libraries and repositories.
This pipeline serves a dual purpose: it streamlines the evaluation process and enhances the comparability of results across different algorithms and datasets. 
The pipeline scheme is shown in Figure~\ref{fig:pipeline_scheme}.

The models evaluated in this study are collaborative filtering methods that use only user-item interaction data. While industrial scenarios often include rich user and item features, we focused exclusively on interaction data. This approach allows for a straightforward comparison of algorithms, including those introduced recently, across a diverse dataset spectrum. Consequently, our choice enabled us to compare numerous algorithms and datasets, making our comparison the most extensive in the current literature.

\subsection{Datasets and Preprocessing} 
In our benchmarking process, we use 30 public datasets, each with timestamps, spanning seven diverse domains.
These datasets cover many business areas, including e-commerce, social networks, and entertainment.
Alongside the utilization of the 28 established public datasets, we introduce two new ones, namely Zvuk and MegaMarket, with their details provided in the Appendix~\ref{appendix:a}. 
This diversity is summarized in Table~\ref{tab:datasets_domains}. 

\begin{table}[!htb]
    \centering
    \begin{tabular}{p{2cm}lp{4.8cm}}
        \hline
        Domain & \# & Datasets \\ 
        \hline
        Movies and clips & 9 &
        Movielense (1, 10, 20M)~\cite{movielens2015}, Netflix~\cite{bennett2007netflix}, 
        Douban movies~\cite{hao:sr}, 
        Amazon TV~\cite{ni2019justifying},
        Kuairec (full/small)~\cite{gao2022kuairec}, Rekko~\cite{rekko} \\
        \rowcolor{gray!10}
        Food and beverage & 5 & BeerAdvocate~\cite{mcauley2012learning}, RateBeer~\cite{ratebeer}, 
        Food~\cite{majumder-etal-2019-generating}, 
        Amazon FineFoods~\cite{ni2019justifying}, Tafeng~\cite{tafeng} \\
        Social networks (SN) & 4 &
        Yelp review~\cite{yelp}, Epinions~\cite{richardson2003trust},  RedditHyperlinks~\cite{kumar2018community}, DianPing~\cite{LiWTM15}  \\
        \rowcolor{gray!10}
        Books & 3 & 
        MTS library~\cite{mtslibrary}, 
        Douban books~\cite{hao:sr}, GoodReads\cite{DBLP:conf/recsys/WanM18, DBLP:conf/acl/WanMNM19} \\
        Location-based SN & 3 & Gowalla~\cite{cho2011friendship}, Brightkite~\cite{cho2011friendship}, FourSquare~\cite{yang2013fine} \\
        \rowcolor{gray!10}
        Music & 3 \textbf{+ 1} &
        Amazon CDs~\cite{ni2019justifying},
        Amazon Musical Instruments~\cite{ni2019justifying},
        Douban music~\cite{hao:sr}, 
        \textbf{Zvuk [new]} \\
        E-commerce &  1 \textbf{+ 1} & Retailrocket~\cite{dkabrowski2021efficient}, \textbf{MegaMarket [new]} \\
        \hline
    \end{tabular}
    \caption{Dataset distribution by domains, \# is the number of datasets in a domain. The newly introduced \emph{Zvuk} and \emph{MegaMarket} expand the most data-scarce domains.}
    \label{tab:datasets_domains}
\end{table}

Implicit feedback-based recommendation systems are increasingly prevalent, primarily due to the frequent absence of explicit rating information in many applications.
Therefore, datasets that initially include item ratings are usually transformed into binary signals, an approach we have also implemented in our evaluation.~\cite{anelli2022top, sun2022daisyrec}.
We have introduced a dataset-specific threshold parameter, denoted as $\tau$, to filter out interactions falling below this threshold. Such interactions are considered negative feedback and are thus removed from the datasets.
For more on determining the optimal $\tau$ value for individual datasets, refer to Appendix~\ref{appendix:a1}.

In their initial state, the datasets exhibit a highly sparse nature, characterized by a substantial proportion of users interacting with a limited number of items, often fewer than five. 
As part of the evaluation process, preprocessing steps are applied to filter out inactive users and items.
Most researchers either adopt a $5$- or $10$-filter/core preprocessing~\cite{sun2020we, sun2022daisyrec}.
$F$-filter and $F$-core filtering techniques differ.
The former simultaneously filters items and users in a single pass, while the latter employs iterative filtering until all users and items have a minimum of $F$ interactions.
We adopt the 5-filter~\footnote{For the newly introduced datasets, Zvuk and MegaMarket, we have implemented a $50$-filter due to their extensive size.} methodology, prioritizing item filtering before user filtering. 
Thus, each user has a minimum of 5 interactions, but some items might have fewer. 


\subsection{Recommendation Models}


Current recommendation frameworks enable the streamlined integration of widely-used baseline models and recently proposed models. We have leveraged existing implementations of well-known algorithms and developed an evaluation pipeline. This pipeline encompasses dataset filtering, data splitting, metrics computation, and hyperparameter optimization.
The frameworks we have used include Implicit~\cite{Implicit}, LightFM~\cite{Kula15}, RecBole~\cite{zhao2021recbole}, and Replay~\cite{replay-rec}. 

Reflecting on recent relevant research in benchmarking~\cite{anelli2022top, sun2022daisyrec}, we have selected the following categories of algorithms for our analysis:
\begin{itemize}
    \item Non-personalized baseline: Random and Popularity-based recommendations (\textbf{Random} and \textbf{MostPop}).
    \item Neighborhood-based model: 
    \textbf{ItemKNN}~\cite{sarwar2001item}.
    \item Matrix factorization models: 
\textbf{LightFM}~\cite{Kula15}, \textbf{ALS}~\cite{hu2008collaborative}, and
    \textbf{BPR}~\cite{rendle2012bpr}.
    \item Linear models: \textbf{SLIM}~\cite{ning2011slim} and \textbf{EASE}~\cite{steck2019embarrassingly}.
    \item Neural models: \textbf{MultiVAE}~\cite{liang2018variational}, \textbf{LightGCN}~\cite{he2020lightGCN}, and 
    \textbf{LightGCL}~\cite{cai2023lightgcl}.
\end{itemize}
While our selection covers many recent approaches, one can add new algorithms from various sources, thereby expanding the scope and capability of the benchmark.

\subsection{Evaluation Settings}

\textbf{\textit{Data splitting.}} A guiding principle for splitting data into training and test subsets is to resemble the deployment conditions closely~\cite{castells2022offline}.
In the top-N recommendation paradigm, the primary challenge is to infer user preferences from past interactions to predict future ones.
Given this, the training data should chronologically precede the test data, acting as the "history" followed by the "future" at a designated time. 
This approach helps in mitigating the risk of data leakage~\cite{ji2023critical}.
Therefore, we adopt the global temporal splitting strategy with an $80/10/10$ training, validation, and test set ratios following~\cite{campos2014time, meng2020exploring, ji2023critical}.
After splitting, we exclude cold-start users and items with no record in the training set.

\textbf{\textit{Negative Sampling.}} In the context of Recommender Systems evaluation, negative sampling involves prediction and evaluation for only a limited set of non-relevant items and known relevant items instead of full item list scoring. 
These non-relevant items are chosen from a candidate item pool.
Although sampling strategies like the Uniform Sampler have been used to avoid biases in evaluating RecSys algorithms and to boost computational efficiency~\cite{bellogin2017statistical, yang2018unbiased, sun2022daisyrec}, studies have questioned their reliability~~\cite{krichene2020sampled}.
Consequently, our evaluation involves testing on all unobserved items.

\textbf{\textit{Evaluation Metrics.}} The precise interpretation of popular quality metrics in the field lacks a consensus, and it is often observed that the more complex a metric is, the greater the scope for varying interpretations~\cite{tamm2021quality}. Therefore, offering a detailed evaluation protocol for reproducibility and clarity in the assessment is crucial.
In light of this, our approach is meticulous: we precisely define each metric and accurately compute them within our established pipeline.

To evaluate the performance of our models, we employ a spectrum of standard quality metrics, such as \textit{Precision@k}, \textit{Recall@k}, \textit{nDCG@k}, \textit{MAP@k}, \textit{HitRate@k}, and \textit{MRR@k}.
Our evaluation scope extends further, incorporating beyond-accuracy objective metrics that provide a more comprehensive view of model effectiveness.
These include \textit{Coverage@k}, \textit{Diversity@k}, and \textit{Novelty@k}~\cite{kaminskas2016diversity}. 

\textbf{\textit{Hyperparameter Tuning.}}
Hyperparameter optimization is crucial for achieving optimal performance of machine learning algorithms and ensuring reliable benchmarking.
The paper~\cite{schnabel2022we} highlights that most RecSys baselines can attain between $90-95\%$ of their maximum performance within the initial $40$ iterations when using Bayesian optimization.
Leveraging this insight, we utilize the Optuna framework~\cite{akiba2019optuna} and apply the Tree of Parzen Estimators (TPE) algorithm for hyperparameter tuning.
In alignment with prior research~\cite{anelli2019discriminative, zhao2022revisiting, sun2020we}, we conduct  hyper-parameter
optimization with respect to \textit{nDCG@10} for each baseline on each dataset.


After determining the optimal hyperparameters, we execute a final training on the consolidated training and validation sets.
This procedure ensures that all available interactions up to the test timestamp are incorporated, including the most recent ones.

\subsection{Metrics Aggregation Methods}

In our benchmarking process, we utilize a matrix containing acquired metrics from various datasets and apply numerous aggregation approaches to analyze these data.

Once evaluation metrics are collected, we should define a method to rank algorithms using performance scores. 
Our pipeline uses well-established methods to aggregate performance to a single rank score over multiple datasets.
These aggregations are adopted from general Machine learning practice and reused for our problem of RecSys methods ranking. 

The list of aggregators includes arithmetic, geometric, and harmonic mean aggregations of a quality metric, CD diagrams~\cite{demvsar2006statistical} emphasizing mean ranks; Dolan-Moré (performance) curves \cite{dolan2002benchmarking} featuring AUC values, and algorithms inspired by the social choice theory, specifically the Copeland, and MinMax rules, proposed for aggregation of results over various NLP tasks~\cite{rofin2022vote}.

\section{EXPERIMENTS AND RESULTS}

\subsection{Metrics}

\begin{figure}[!h]
    \centering
    \includegraphics[width=0.75\columnwidth]{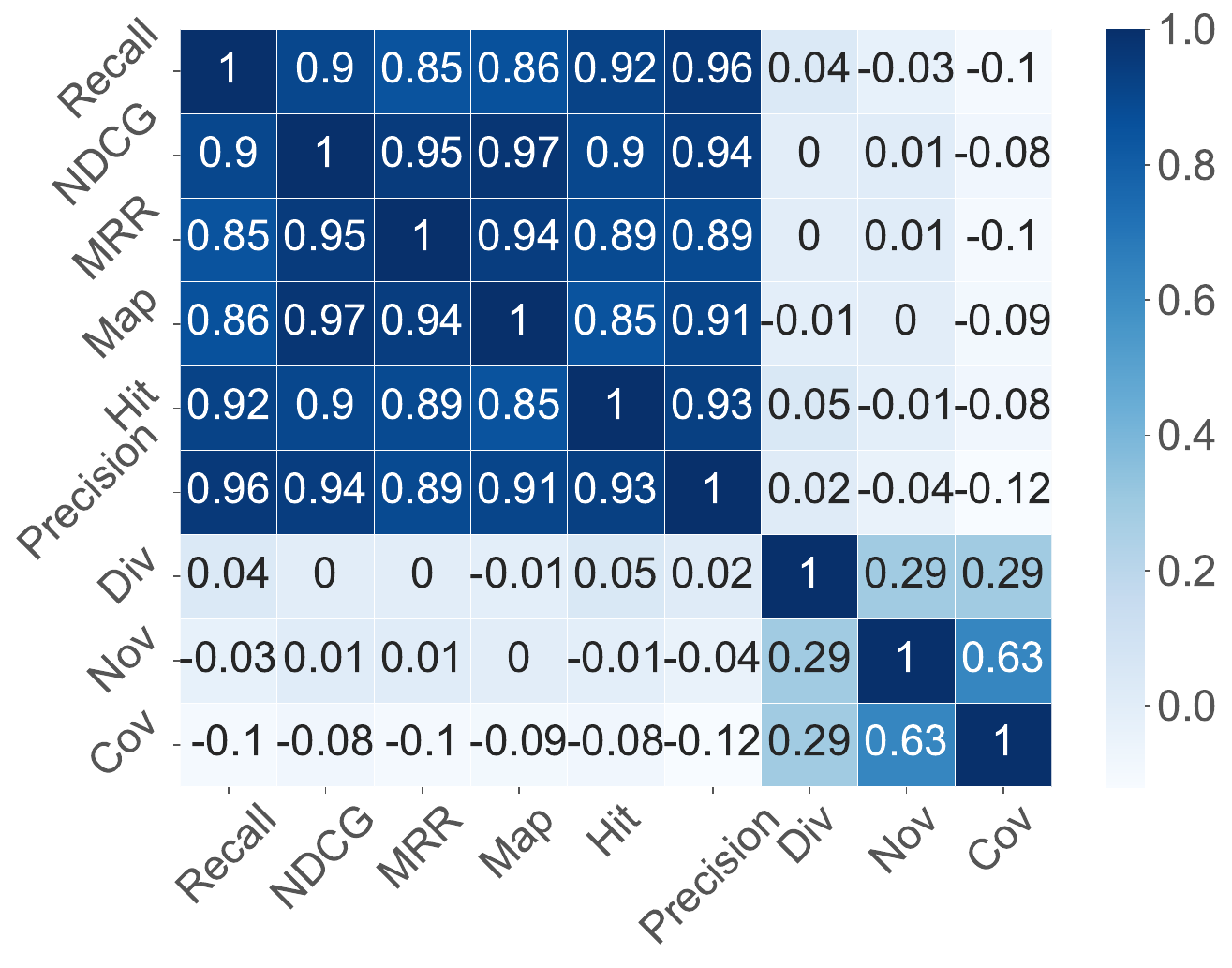}
    \caption{Spearman correlation between metrics for $k=10$. Darker blue indicates stronger correlations.}
    \label{fig:gm_metrics}
\end{figure}

Our experiments begin with the collection of performance metrics to evaluate 11 recommendation algorithms across 30 datasets. These metrics include User Preference Accuracy, Ranking Quality, and Beyond-Accuracy metrics. Following the data collection, we perform an initial analysis of the accumulated results, focusing on Spearman's correlation for each dataset and subsequently computing the average correlation scores. These results are consolidated into a correlation heatmap, illustrating the relationships among all pairs of metrics, as shown in Figure~\ref{fig:gm_metrics}.

The structure of our correlation matrix is similar to the smaller-scale experiment conducted by Zhao et al.~\cite{zhao2022revisiting}. The heatmap reveals that the Accuracy and Ranking metrics located in the top left corner ($Recall$, $nDCG$, $MRR$, $MAP$, $HitRate$, $Precision$) exhibit high correlations with each other, surpassing $0.8$. In contrast, the Beyond-Accuracy metrics positioned at the bottom demonstrate weak correlations with the rest. This distinction can be attributed to the different goals of these metrics, as Beyond-Accuracy metrics do not straightforwardly describe recommendation quality.

Moreover, our findings indicate that nDCG has the highest correlation ($\geq 0.9$) with accuracy and ranking metrics. This observation reinforces the usage of $nDCG$ in benchmarking and during hyper-parameter optimization as an optimization objective. Consequently, in the subsequent experiments, we utilize $nDCG@10$ unless otherwise stated.

\subsection{Comparative Analysis of Metrics Aggregation Methods}
In this section, our primary focus is on exploring various methods to aggregate metrics derived from diverse datasets.
We describe the considered aggregation approaches and then analyze the ranking of RecSys models using these methods.

Given the RecSys specifics~\cite{zhu2022bars,said2014comparative}, we identify key requirements for a ranking method:
\begin{itemize}
    \item \textbf{Ranking:} The benchmarking system should rank methods according to their performance.
    \item \textbf{Metric Value Consideration:} It should consider the metric values and their relative differences for specific problems, not just the relative positions.
    \item \textbf{Interpretability:} The results should be interpretable, providing clear insights into model comparisons.
    \item \textbf{Significance determination:} The system should explicitly define the significance of performance differences.
    \item \textbf{Agnosticism to adversarial manipulations:} The ranking should be robust to malicious influence.
\end{itemize}

\subsubsection{Considered aggregation approaches}

We consider the following ways to aggregate metrics:

\paragraph{Mean Ranks (MR)}
MR is used in Critical Difference diagrams and computes the average ranks of methods across all datasets.

\paragraph{Mean Aggregations}
These approaches utilize classic mathematical averages, such as \textit{Mean aggregation (MA)}, \textit{Geometric mean (Geom. mean)} and \textit{Harmonic mean (HM)}, calculating across datasets for each model as a single model score.

\paragraph{Dolan-Moré Area Under Curve (DM-AUC)}
This relates to the Dolan-Moré performance profiles defined for $\hat{\beta} \geq \beta \geq 1$ as
\[
    p_i(\beta) = \frac{1}{d}\left|\left\{t: \beta q_{ti} \geq \max_j q_{tj} \right\}\right|, 
\]
where $i$ is the index of the curve that corresponds to a RecSys model, $t$ is the problem index, $d$ is the number of datasets, $\hat{\beta}$ is a hyperparameter limiting the values of $\beta$, $q_{ti}$ is the metric value that corresponds to a RecSys model $i$ and the problem $t$, and $q_{tj}$ is the metric value that corresponds to a RecSys model $j$ and the problem $t$.
The DM curve for a specific $\beta$ reports the number of problems for which the model performs no more than $\beta$ times worse than the best model for that problem (e.g., $p_i(1)$ represents the share of problems where the $i$-th algorithm is the best). 

In the case of aggregation, we take the area under the DM curve divided by the sum of all areas. In our experiments, we fixed $\hat{\beta} = 3$; below, we show that ranking remains stable across a wide range of $\beta$ values.

\paragraph{Dolan-Moré leave-best-out (DM LBO)}
This approach ranks algorithms by performing the following steps:
\begin{enumerate}
    \item Calculate DM AUC (area under curve) scores;
    \item Choose the best method using DM AUC and remove it;
    \item The method that is removed is assigned a rank based on the iteration in which it was dropped;
    \item Repeat the previous steps using the remaining methods.
\end{enumerate}

\paragraph{Social Choice Theory}
The last two aggregating approaches considered, \textit{Copeland} and \textit{Minimax}, are majority-relation-based rules. 
A majority relation for two methods $m_A$, $m_B$ ($m_A \succ m_B$) holds, if $m_A$ has a higher metric value than $m_B$.  
\textit{Copeland} method defines the aggregation score $u(m)$ as $u(m_A) = |L(m_A)| - |U(m_A)|$,
where $L(m_A) = \{m | m_A \succ m\}, U(m_A) = \{m | m \succ m_A\}$.
The \textit{Minimax} uses a score $s(m_A, m_B)$, representing the number of datasets for which method $m_A$ has a higher score than $m_B$. The aggregated score is given by
$ u(m_A) = - \max_B s(m_B, m_A)$.

\begin{table*}[!htb]
    \centering
    \resizebox{\textwidth}{!}{
    \begin{tabular}{lllllllll}
    \toprule
    {Ranking} &          DM AUC $\uparrow$ &       DM LBO $\downarrow$ &       Mean ranks $\downarrow$ &               MA $\downarrow$ &       Geom. mean $\downarrow$ &       Harm. mean $\downarrow$ &       Copeland $\uparrow$ &          Minimax $\uparrow$ \\
    position &                 &              &                  &                  &                  &                  &                &                  \\
    \midrule
    1              &  EASE: 0.121 &  EASE: 1 &  EASE: 2.933 &  EASE: 0.069 &  EASE: 0.042 &  EASE: 0.023 &  EASE: 10.0 &  EASE: -0.0 \\
    2              &  LightGCN: 0.111 &  LightGCN: 2 &  MultiVAE: 4.107 &  LightGCL: 0.065 &  LightGCN: 0.038 &  LightGCN: 0.021 &  MultiVAE: 8.0 &  SLIM: -21.0 \\
    3              &  MultiVAE: 0.111 &  LightGCL: 3 &  LightGCN: 4.363 &  LightGCN: 0.064 &  LightGCL: 0.038 &  ALS: 0.02 &  LightGCN: 6.0 &  MultiVAE: -22.0 \\
    4              &  LightGCL: 0.11 &  MultiVAE: 4 &  SLIM: 5.247 &  MultiVAE: 0.061 &  MultiVAE: 0.038 &  LightGCL: 0.02 &  SLIM: 3.0 &  LightGCN: -22.0 \\
    5              &  ALS: 0.106 &  ALS: 5 &  ALS: 5.32 &  LightFM: 0.059 &  ALS: 0.035 &  MultiVAE: 0.02 &  ALS: 2.0 &  LightGCL: -23.0 \\
    6              &  ItemKNN: 0.1 &  ItemKNN: 6 &  LightGCL: 5.5 &  SLIM: 0.058 &  LightFM: 0.034 &  ItemKNN: 0.018 &  LightGCL: 0.0 &  ALS: -24.0 \\
    7              &  LightFM: 0.1 &  LightFM: 7 &  LightFM: 5.707 &  BPR: 0.057 &  ItemKNN: 0.033 &  LightFM: 0.017 &  LightFM: -1.0 &  BPR: -25.0 \\
    8              &  SLIM: 0.093 &  BPR: 8 &  ItemKNN: 6.25 &  ALS: 0.057 &  BPR: 0.03 &  BPR: 0.014 &  ItemKNN: -4.0 &  ItemKNN: -26.0 \\
    9              &  BPR: 0.088 &  SLIM: 9 &  BPR: 6.793 &  ItemKNN: 0.056 &  SLIM: 0.025 &  MostPop: 0.006 &  BPR: -6.0 &  LightFM: -26.0 \\
    10             &  MostPop: 0.058 &  MostPop: 10 &  MostPop: 9.067 &  MostPop: 0.041 &  MostPop: 0.017 &  SLIM: 0.003 &  MostPop: -8.0 &  MostPop: -29.0 \\
    11             &  Random: 0.003 &  Random: 11 &  Random: 10.8 &  Random: 0.007 &  Random: 0.001 &  Random: 0.0 &  Random: -10.0 &  Random: -30.0 \\
    \bottomrule
    \end{tabular}}
    \caption{Rankings of RecSys methods according to different aggregation approaches with their respective scores. The leaderboard is based on nDCG@10 values.
    }
    \label{tab:aggregation_results}
\end{table*}

\begin{table*}
    \resizebox{\textwidth}{!}{
    \begin{tabular}{lrrrrrrrr}
    \toprule
     &  DM AUC &  DM LBO &  Mean ranks &     MA &  Geom. mean &  Harm. mean &  Copeland &  Minimax \\
    \midrule
    Pareto efficacy & $+$ & $+$ & $+$ & $+$ & $+$ & $+$ & $+$ & $+$ \\
    Using small number of datasets & $+$ & $+$ & $+$ & $-$ & $+$ & $-$ & $+$ & $-$ \\
    Using small number of methods & $+$ & $+$ & $-$ & $+$ & $+$ & $+$ & $+$ & $-$ \\
    Adding a new similar method & $+$ & $+$ & $-$ & $+$ & $+$ & $+$ & $-$ & $-$ \\
    Adding a new best method & $-$ & $+$ & $+$ & $+$ & $+$ & $+$ & $+$ & $-$ \\
    Changing hyperparameters & $-$ & $-$ & NA & NA & NA & NA & NA & NA \\
    Spearman's correlation, $5$ datasets & 0.799 & 0.785	 & 0.825 & 0.717	 & 0.834	 & 0.756	 & 0.816	 & 0.525 \\
    Spearman's correlation, $10$ datasets & 0.895 & 0.887 & 0.912 & 0.825	 & 0.899 & 0.885	 & 0.907	 & 0.767 \\

    \bottomrule
    \end{tabular}
    }
    \caption{General results of rankings reliability. $+$ stands for the availability of a feature, $-$ stands for an absence, and NA stands for not applicable.
    }
    \label{tab:aggregation_other_results}
\end{table*}



\subsubsection{Comparison of RecSys algorithms}

\begin{figure}[!ht]
    \centering
    \includegraphics[width=0.72\columnwidth]{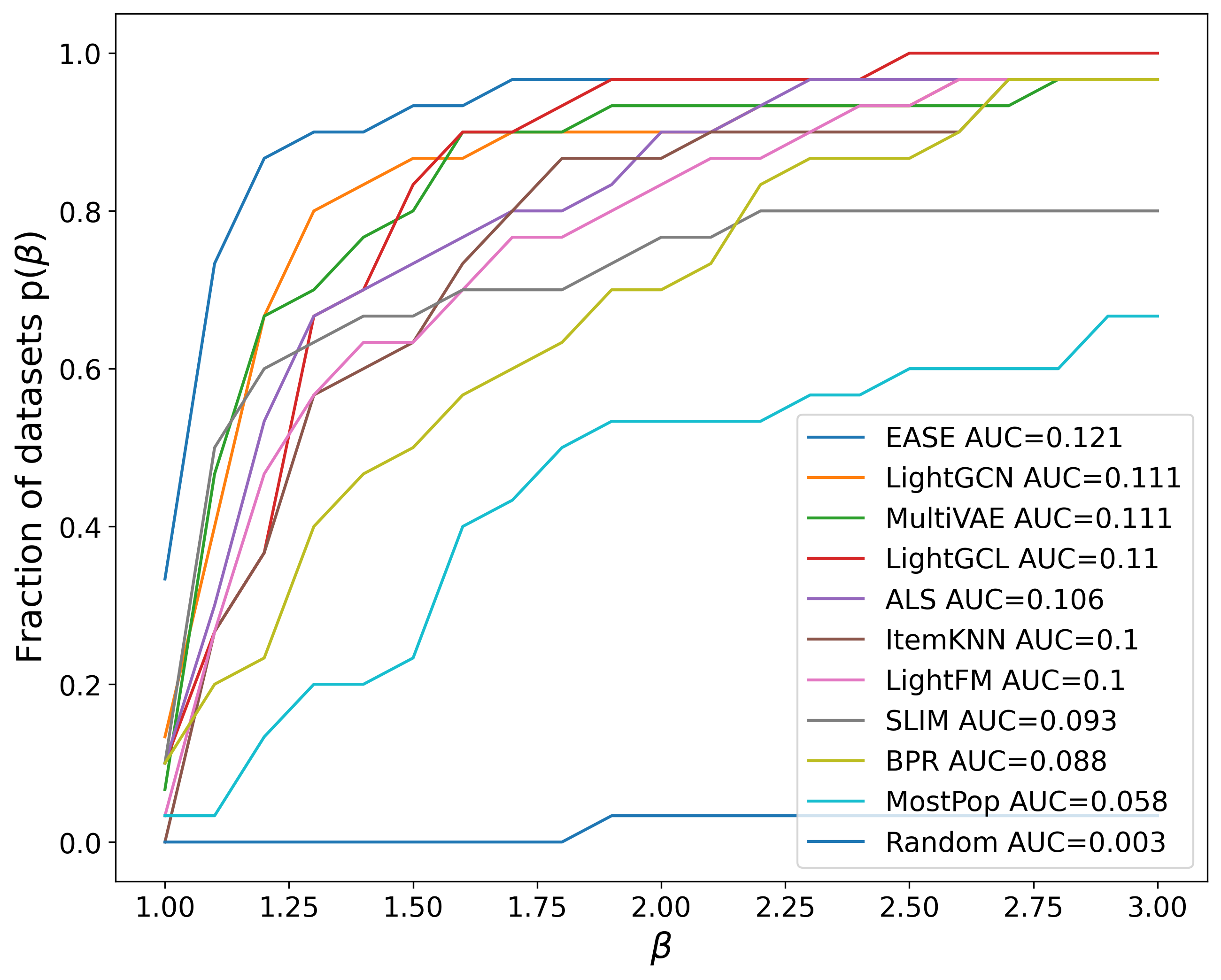}
    \caption{Performance profiles for the comparison of RecSys algorithms. The higher the curve, the better the performance of the algorithm. We also provide AUCs for each approach.}
    \label{fig:dolan_more_curve}
\end{figure}

\begin{figure}[!h]
    \centering
    \includegraphics[width=0.9\columnwidth]{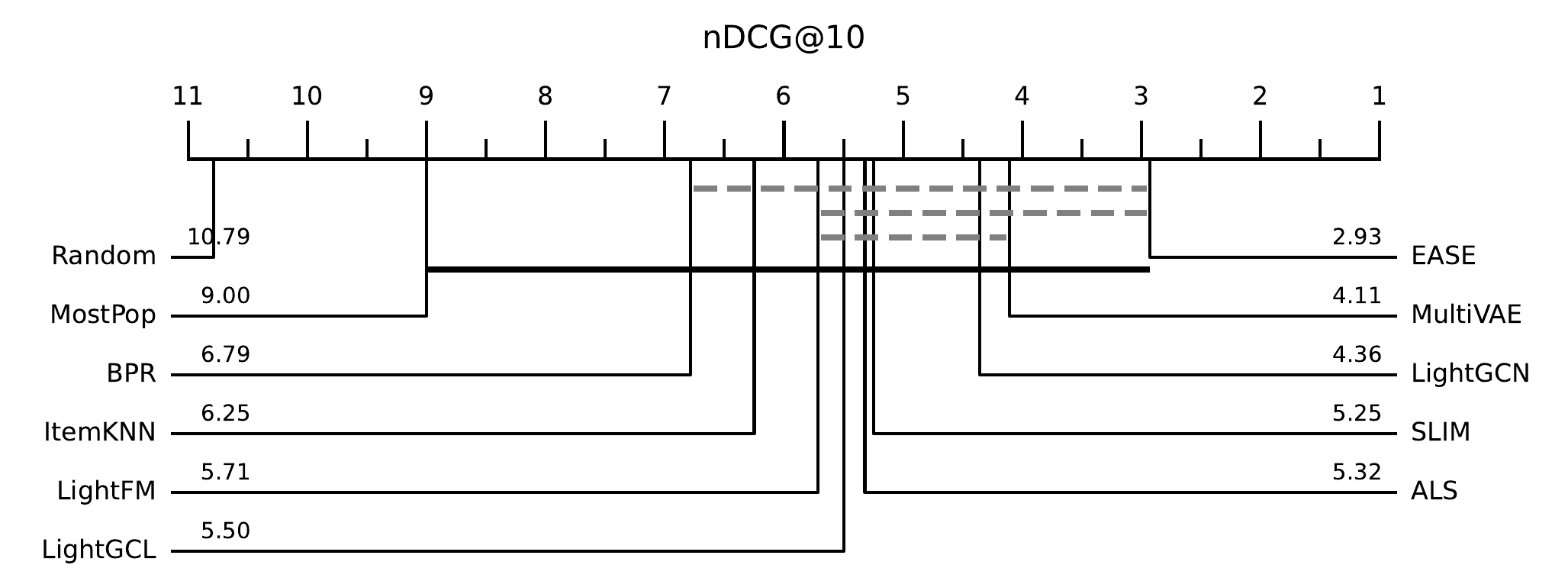}
    \caption{The Critical Difference diagram for the comparison of RecSys algorithms. The numbers represent the mean ranks of methods over all datasets. Thick horizontal lines represent a non-significance based on the Wilcoxon-Holmes test, while dashed horizontal lines represent non-significance according to the Bayesian Signed-Rank test.}
    \label{fig:cd_diagram}
\end{figure}

One of our objectives is to present interpretable results that facilitate swift visual comparisons of the performance of RecSys algorithms across multiple datasets.
We present these visual comparisons using DM performance profiles and a CD diagram.
The DM profiles are in Figure~\ref{fig:dolan_more_curve}.
Further, we use the presented DM AUC to rank the algorithms.
The CD diagram can be found in Figure~\ref{fig:cd_diagram}.
In addition to the traditional CD diagram that includes the pairwise Wilcoxon test, we have introduced the Bayesian Signed-Rank test, indicated by dashed horizontal grey lines. 
We exclude the concept of ROPE from our analysis because it requires homogeneity among the set of metrics, which is not applicable in RecSys. 
This inhomogeneity also leads
to the absence of statistical significance in the CD.
While we use a large number of datasets, due to their diversity, the ranks of approaches change a lot.

However, our findings indicate that EASE emerges as the winner for both options.
There is no distinct second-place algorithm, as areas under the performance profiles for LightGCN and MultiVAE are almost identical. 
Finally, all methods perform significantly better than a random approach and are mostly superior to the MostPop baseline.


\paragraph{Leaderboard for different aggregations}

Different aggregation methods yield distinct rankings for the approaches considered. 
With the set of $30$ datasets, the ranks presented in Table~\ref{tab:aggregation_results} consistently identify EASE as the top-performing approach.
For the subsequent top positions, we have a pair of candidates for most aggregations: MultiVAE and LightGCN.




\subsubsection{Comparison of reliability of aggregations}

To ensure a reliable aggregation method for benchmarking, it should demonstrate stability under various perturbations, including adversarial ones.
We conduct an analysis to assess the robustness of the presented aggregation methods.

First, we determine rankings based on 30 datasets across all methods, establishing them as our reference benchmarks.
Next, we examine the sensitivity of these rankings to the following modifications of the input matrix of quality metrics:
\begin{enumerate}
    \item Inclusion or exclusion of a dataset.
    \item Introduction or removal of a RecSys algorithm.
    \item Incorporation or exclusion of a slightly superior/inferior method to a particular algorithm, exploring all possible permutations.
    \item Adjustments in the hyperparameters of an aggregation method.
\end{enumerate}

\paragraph{Change of the set of used datasets} 
We measure the correlation between the final rankings and the references using the Spearman correlation coefficient $\rho$. 

\begin{figure}[!t]
    \centering
    \includegraphics[width=0.9\columnwidth]{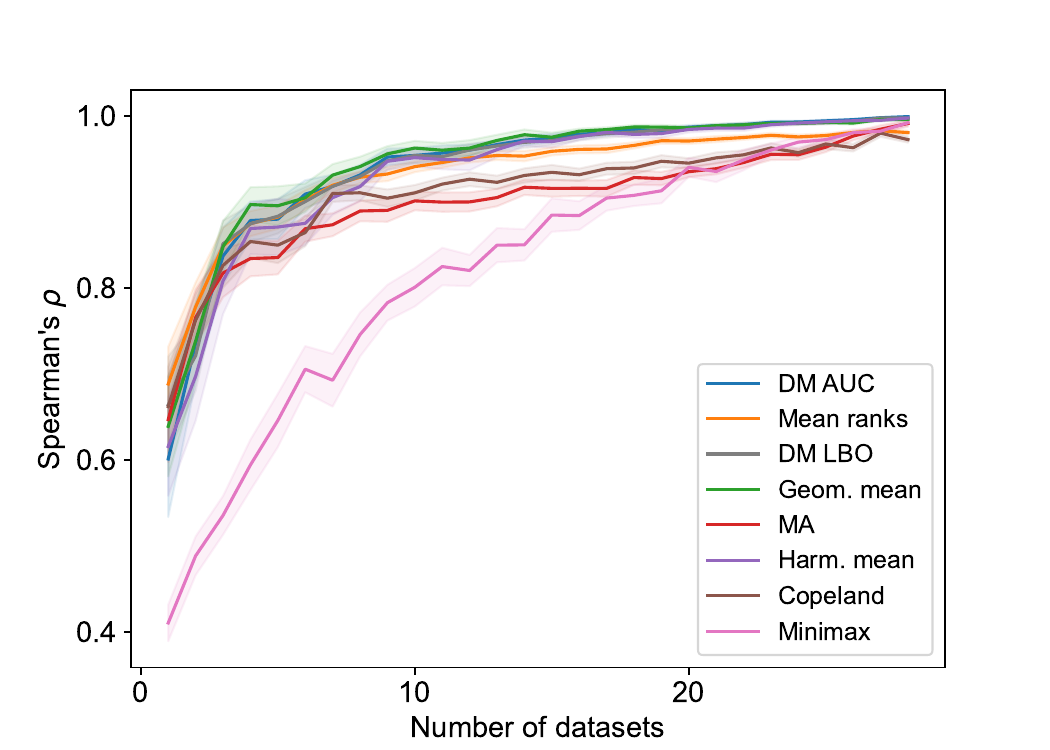}
    \caption{Stability of aggregations with respect to the number of used datasets.}
    \label{fig:drop_data_case}
\end{figure}

The results for the case of dropping datasets are presented in Figure~\ref{fig:drop_data_case}. 
All aggregations exhibit a relatively stable behaviour, except for Minimax, which shows a low $\rho$ after dropping 15 datasets. 
On the contrary, the best-performing method is Geom. mean, Harm. mean and DM LBO, with their average metric values being less influenced by specific datasets. Overall, different aggregation techniques tend to produce similar rankings if the number of datasets is large enough.

Furthermore, we explore the case when we use only five datasets to calculate ranks.
We randomly sampled 100 pairs of subsets of size five and calculated Spearman's correlation between aggregations for each pair of sets of datasets.
The results are in Table~\ref{tab:aggregation_other_results}.
Aggregations are less stable in this case.
Moreover, \emph{MA}, \emph{Harm. mean} and \emph{Minimax} methods have Spearman's correlation of less than $0.8$.
As in the case of dropping datasets, the outlier is \emph{Minimax} model with low $\rho$.
The \emph{Mean ranks} and \emph{Copeland} methods perform the best in this scenario, demonstrating equivalent $\rho$ values.

\paragraph{Elimination of RecSys methods}
In this scenario, we compute Spearman's correlation between the results for all methods and the results with the exclusion of some methods.

\begin{figure}[!t]
    \centering
    \includegraphics[width=0.9\columnwidth]{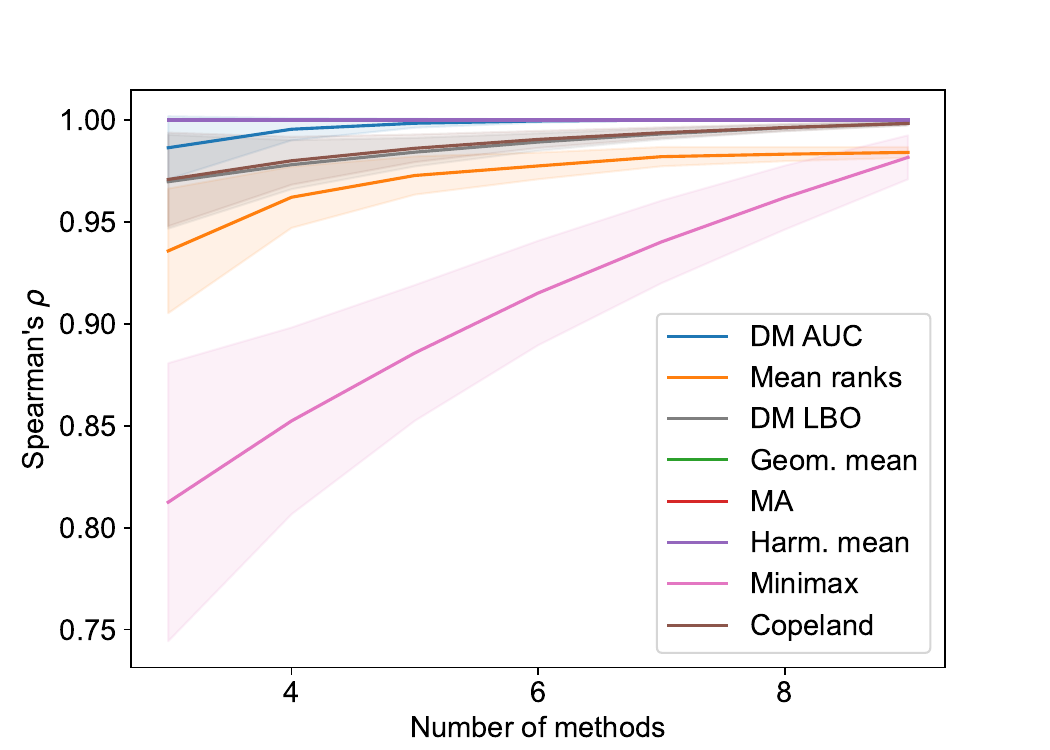}
    \caption{Stability of aggregations with respect to the number of used methods.}
    \label{fig:drop_method_case}
\end{figure}

The results are presented in Figure~\ref{fig:drop_method_case}.
We observe that most aggregating methods exhibit relatively stable behaviour, except for \emph{Minimax}.
As the number of discarded methods increases, the likelihood of changing the best method also increases.
This notably impacts the \emph{Mean ranks}, \emph{DM AUC} and \emph{Copeland} methods.

\paragraph{Changing the hyperparameters of the aggregating method}
In the paper, only the \emph{DM AUC} and \emph{DM LBO} aggregating methods have adjustable hyperparameters ($\hat{\beta}$ - the maximum value of the ratio of the best metric value and the one under consideration, the right boundary of the X-axis in the performance profiles). 
We calculate the Spearman correlation between the case when $\hat{\beta} = 3$ and the case when $\hat{\beta}$ can take any value. 
The results presented in Figure~\ref{fig:hp_correlation_results} demonstrate that $\hat{\beta}$ can influence the rankings, underscoring the significance of determining and fixing the optimal value. The ranking remains stable over a wide range of $\hat{\beta}$ values, with \emph{DM LBO} showing more robustness compared to the pure \emph{DM AUC}.

\paragraph{Additional experiments.}
In Appendix~\ref{appendix:c}, we explore the stability of a ranking provided by aggregations after the inclusion of a new method slightly superior or inferior to an existing one.
There, all methods, except for \emph{Mean ranks}, demonstrate stability under adversarial perturbations.
Moreover, our analysis delves into specific intrinsic properties of a ranking system, such as Pareto efficiency when using a small number of datasets. 
An aggregation method is considered Pareto efficient if it outperforms another method for all metrics. 
The results of all tests are presented in Table~\ref{tab:aggregation_other_results}.

\subsection{Dataset Characteristics}

In addition to the performance benchmark, our study explores the connection between specific dataset characteristics and recommendation quality.
We utilize user-item interaction matrix properties from~\cite{deldjoo2021explaining}. 
These properties serve as problem characteristics and encompass various aspects, including the size, shape, and density of the dataset (\emph{SpaceSize, Shape, Density}), as well as counts of users, items, and interactions (\emph{Nu, Ni, Nr}). 
We also consider interaction frequencies per user and item (\emph{Rpu, Rpi}), Gini coefficients that describe interaction distribution among users and items (\emph{Giniu, Ginii}), and statistics related to popularity bias and long-tail items (\emph{APB, StPB, SkPB, KuPB, LTavg, LTstd, LTsk, LTku})~\cite{deldjoo2021explaining}. 
These characteristics of the 30 selected datasets exhibit a wide range of variability.

\begin{figure}[h]
    \centering
    \includegraphics[width=0.99\columnwidth]{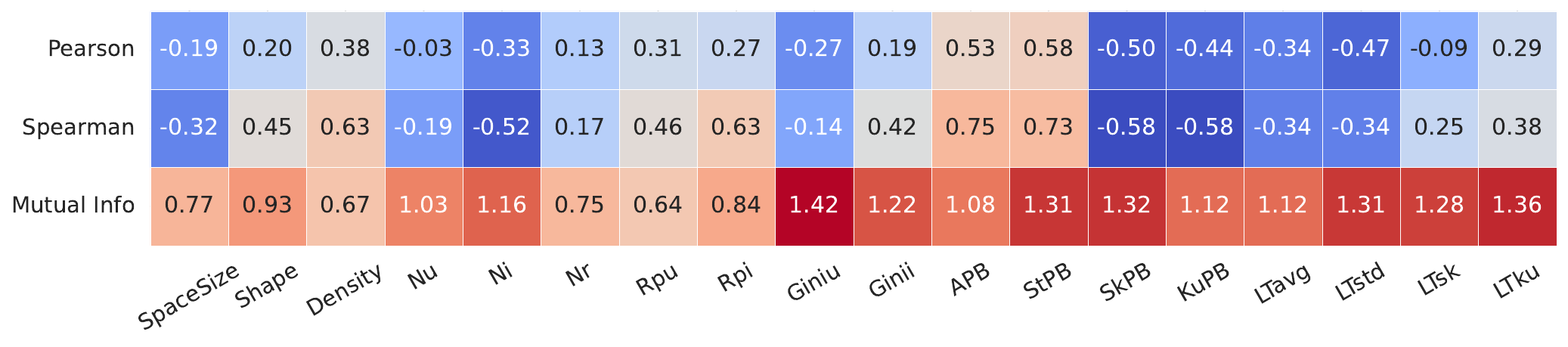}
    \caption{Pearson, Spearman correlations, and Mutual information between data characteristics and RecSys algorithms performance.}
    \label{fig:correlations}
\end{figure}

To establish a connection between these data characteristics and our primary quality metric, $nDCG@10$, we employed three distinct measures: Pearson product-moment correlation, Spearman rank --- an order correlation, and Mutual information --- a nonlinear alternative. The obtained values are in Figure~\ref{fig:correlations}. 

Datasets exhibiting higher levels of popularity bias \emph{APB} and \emph{Density} tend to simplify the prediction tasks for recommender models. 
Conversely, datasets exhibit long-tailed item distributions, increased item diversity, and pronounced Popularity Bias, presenting a greater challenge for recommender models. 
Furthermore, the moderate mutual information values emphasize the practical impact of these characteristics on the performance of the models. 

\subsection{Optimizing Dataset Selection for Benchmarking}

Using 30 public datasets provides diverse evaluation characteristics but is not computationally efficient. To decrease the use of datasets while preserving variability, it would be practical to select datasets that may belong to the same group. We employ the KMeans approach to split datasets into multiple clusters, using data characteristics from the previous section as feature representations.



\textbf{\textit{Principal datasets selection.}} To decrease time consumption and minimize the degradation of benchmarking, we can run it only for a limited number of datasets, carefully selecting them. We use several approaches for selection: Random, KMeans, A-optimality, and D-optimality approaches.

In Random, we uniformly at random select a subset of datasets from the set.
The KMeans identifies core datasets as the closest ones to cluster centers for clusters being selected in the space of data characteristics~\cite{chin2022datasets}.
Two additional baselines are A-optimality and D-optimality.
They constitute two fundamental criteria focused on obtaining the lowest possible error of a model that predicts performance and the error for parameters estimation of a model~\cite{pukelsheim2006optimal}.
Technical details are provided in Appendix~\ref{appendix:optimal_design}. 



By selecting six datasets per method and calculating Spearman’s correlation across 500 simulations, our results indicate superior performance of the KMeans, as shown in Table~\ref{tab:principal_datasets}. 


\begin{table}[!h]
    \centering
    \begin{tabular}{llll}
    \toprule
    Method  &       nDCG\@10 &       HitRate\@10 &               Coverage  \\
    \midrule
    Random & 0.834 & 0.855 & 0.939 \\
    D optimal & 0.805 & 0.819 & 0.913 \\
    A optimal & 0.669 & 0.687 & 0.887 \\
    KMeans & \textbf{0.845} & \textbf{0.900} & \textbf{0.982} \\
    \bottomrule
    \end{tabular}
    \caption{Spearman correlation among metrics across six selected datasets compared to the entire set of 30 datasets.}
    \label{tab:principal_datasets}
\end{table}

\begin{figure}[!t]
    \centering
    \includegraphics[width=0.9\columnwidth]{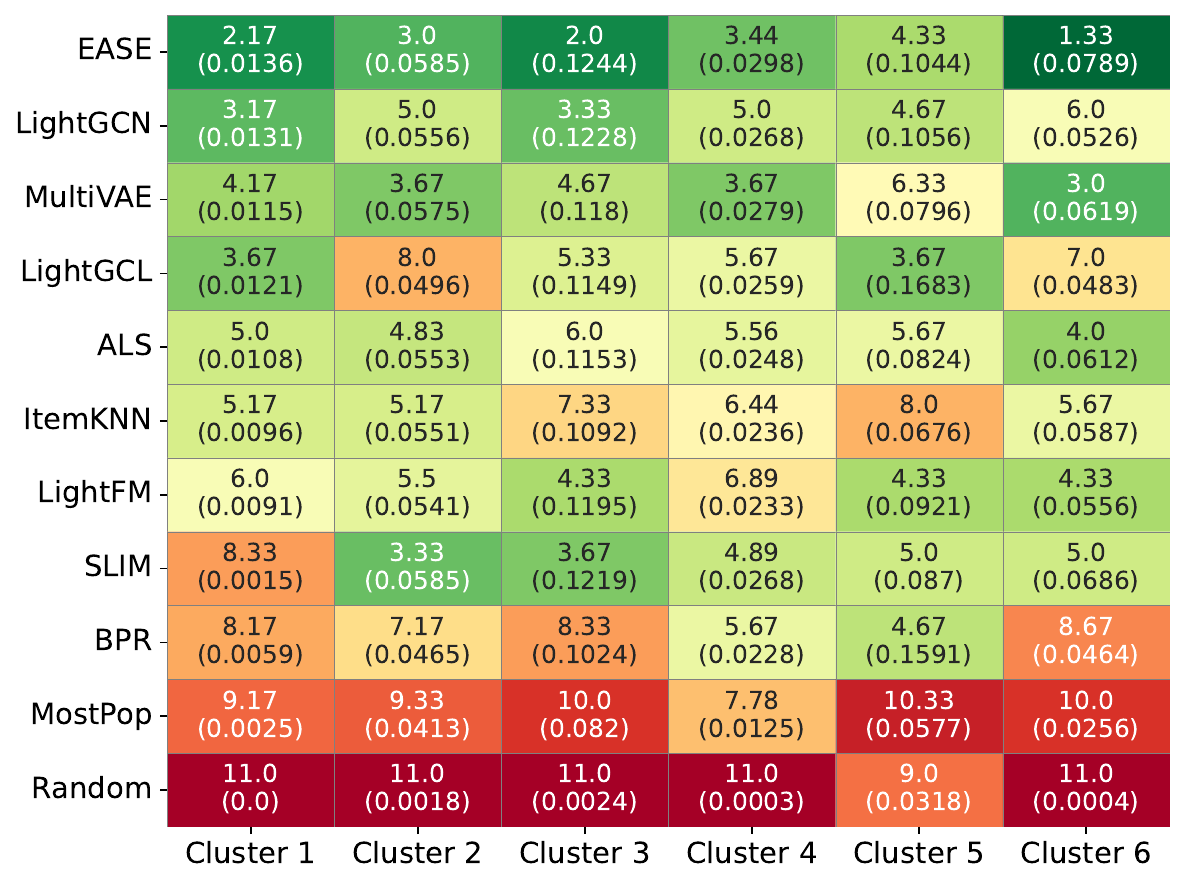}
    \caption{Ranks and Geom. mean (in brackets) aggregated $nDCG@10$ on different clusters of datasets.}
    \label{fig:clusters_ranks}
\end{figure}

\textbf{\textit{Clustering datasets using their characteristics}}
With the clustering approach described above, we have generated rankings and metrics for different clusters as illustrated in Figure~\ref{fig:clusters_ranks}, and the specific datasets for each cluster are listed in Table~\ref{tab:datasets_clusters}.

 \begin{table}[!b]
    \centering
    \begin{tabular}{ll}
    \toprule
    Datasets  &       Cluster   \\
    \midrule
    Amazon CDs, Amazon TV, Gowalla, MTS library,  & 1 \\
    Amazon Musical Instruments, Yelp review & \\
    \rowcolor{gray!10}
    BeerAdvocate, DianPing, Douban movies,  & 2 \\
    \rowcolor{gray!10}
    Movielens 1M, RedditHyperlinks, Rekko & \\
    Movielens 10M, Movielens 20M, Ratebeer & 3 \\
    \rowcolor{gray!10}
    Amazon FineFoods, Brightkite, Douban Books, & \\
    \rowcolor{gray!10}
    Douban music, Epinions, Food, & 4 \\
    \rowcolor{gray!10}
    GoodReads, Retailrocket, Tafeng & \\
    Foursquare, Kuairec full, Kuairec small & 5 \\
    \rowcolor{gray!10}
    Netflix, MegaMarket, Zvuk & 6 \\
    \rowcolor{gray!10}
    \bottomrule
    \end{tabular}
    \caption{Datasets clusters obtained by described approach.}
    \label{tab:datasets_clusters}
\end{table}

\textbf{Cluster 1} consists of six datasets characterized by a high number of items relative to users. For example, the Amazon TV dataset includes approximately $\sim50$K users, $\sim216$K items, and $\sim2$M interactions. \textbf{Cluster 2} also includes six datasets, each marked by moderate characteristic values and relative sparsity. For instance, the Reddit dataset comprises $\sim4$K users and items each, with around $\sim70$K interactions. \textbf{Cluster 3} is smaller, containing just $3$ datasets, where the number of users significantly exceeds the number of items. This pattern suggests a different interaction dynamic compared to other clusters. \textbf{Cluster 4} includes $9$ datasets with a moderate number of items and users. Typically, the number of items surpasses the number of users in these datasets, which have fewer interactions (generally $\leq$ 1M, with one exception).  Moreover, the SkPB and KuPB are highest in this cluster, meaning items have an imbalanced probability distribution. For instance, the Retail dataset features about $\sim32$K users, $\sim52$K items, and $\sim342$K interactions, with SkPB $\sim2$ and KuPB $\sim6$.
\textbf{Cluster 5} contains only $3$ datasets, which are the smallest in terms of user and item counts but are the most densely populated, with densities ($\sim0.1$). For example, the Kuairec small dataset has about $\sim1400$ users, $\sim3100$ items, and a high density of 0.8. This cluster shows the lowest SkPB, indicating a more uniform item usage across users.
\textbf{Cluster 6} consists of the largest datasets, where both the user and item counts typically $\geq 100K$, and interactions $\geq 20M$, denoting sparse interactions. A prominent dataset in this cluster is MegaMarket, with $\sim184$K users, $\sim167$K items, and $\sim25$M interactions. Note that the characteristics of the datasets and the clusters identified herein are derived from the preprocessed versions of these datasets.

\textbf{\textit{Cluster Analysis Summary.}} The EASE consistently performs well across most clusters, with LightGCL only outperforming it in Cluster $5$. Other approaches exhibit less stability. For example, LightGCL has an average rank of $8$ for Cluster $2$, while SLIM has the second-best average rank for it. This variability is partly due to the differences in cluster complexity, as evidenced by EASE's fluctuating geometric mean from $0.0136$ to $0.1244$. These results underline the importance of using datasets with similar characteristics for effective offline evaluation for business use cases.

\subsection{RecSys Performance Variability over Datasets}
Drawing upon the insights from the benchmarks, we conclude the main part of the article with an analysis of how different algorithms perform in relation to specific dataset characteristics.


\textbf{ItemKNN:} Despite its interpretability and suitability as a baseline, ItemKNN shows limited effectiveness, excelling only in datasets with specific characteristics such as low Shape values or moderate Density, including Movielens 1M, GoodReads, and Amazon MI. These findings align with challenges of sparsity and scalability typical of neighborhood-based models \cite{sarwar2001item}. 



\textbf{Matrix Factorization Methods (LightFM, ALS, BPR):} These algorithms vary in performance depending on dataset characteristics. Our analysis identifies that sparsity remains a significant limitation~\cite{Kula15,hu2008collaborative}. 
While these methods serve as reliable baselines, their efficacy is more pronounced in smaller datasets, like Foursquare, where the data structure may not be extensively sparse. 
LightFM tends to perform less efficiently in terms of computation time in larger datasets. 
ALS demonstrates more robust performance across various scales, making it a versatile choice for both moderately sized and large datasets.




\textbf{Linear Models (SLIM, EASE):}  These models show consistently high performance across a variety of datasets. 
EASE, in particular, excels in datasets with extensive user interaction data, supporting findings from \cite{steck2019embarrassingly}. However, the significant computational demands of EASE, including both time and memory resources, may limit its use in settings with restricted computational capabilities.



\textbf{Neural and Graph-Based Models:} Demonstrating superior performance, especially in dense or highly connected environments, MultiVAE and graph-based method LightGCN are particularly effective. MultiVAE performs well in datasets with moderate user interactions, benefiting from Bayesian priors that help manage the inherent uncertainty in sparse data. Graph-based models excel in datasets with high connectivity, such as social networks (e.g., Gowalla, Yelp), where the relational structure can be fully exploited to enhance recommendation quality.

\section{Conclusions}

Our paper introduces a novel benchmarking system for recommender systems. 
It integrates a rigorous pipeline that leverages multiple datasets, hyperparameter tuning, and validation strategy, as well as an aggregation procedure for metrics across different datasets.
Our approach is interpretable and robust, working for distinct metrics used for RecSys evaluation.
Among the considered methods, EASE is a clear winner with respect to all considered aggregation strategies. 
Other methods show inferior performance on average while being interesting for particular subdomains identified by our clustering scheme.

Further research provides deeper insight related to the stability and efficiency of ranking.
Due to the usage of 30 datasets, two of which are open-sourced in this study, the results are robust in diverse considered scenarios.
Via our clustering procedure, we obtain a collection of $6$ datasets that also provides a consistent ranking, achieving efficiency and reliability simultaneously.
Additional experiments confirm the stability of our benchmark with respect to reducing the number of considered datasets, methods, and adversarial manipulation of the list of methods.
Overall, our research offers a streamlined guide and valuable datasets for advancing recommender system studies that can be used both by practitioners during the selection of a method and researchers during the evaluation of a novel idea.

\section{Acknowledgments}

The work was supported by the Analytical center under the RF Government (subsidy agreement 000000D730321P5Q0002, Grant No. 70-2021-00145 02.11.2021).

\bibliographystyle{ACM-Reference-Format}
\bibliography{articles}

\appendix

%




\section{New Datasets}
\label{appendix:a}
We introduce two new datasets named MegaMarket and Zvuk.
Their size exceeds the sizes of many publicly accessible datasets. Additionally, these datasets are from fields that are less commonly represented in existing research.

\textbf{MegaMarket} documents user interactions, tracking events like views, favorites, additions to carts, and purchases over a five-month period from January 15 to May 15, 2023. It comprises a total of 196,644,020 events involving 3,562,321 items across 10,001 distinct categories, contributed by 2,730,776 unique users.

\textbf{Content}:
\begin{itemize}
    \item \textbf{User IDs}: Distinct identifiers for users, totaling 2,730,776.
    \item \textbf{Datetimes}: Timestamp range from 01.15.2023 00:00:00.708 to 14.05.2023 20:59:59.000.
    \item \textbf{Events}: Categorized by unique codes, comprised of:
    \begin{itemize}
        \item 0: View
        \item 1: Favorites
        \item 2: Add to cart
        \item 3: Purchase
    \end{itemize}
    \item \textbf{Item IDs}: Specific identifiers for items, amounting to 3,562,321.
    \item \textbf{Category IDs}: Denoting which of the 10,001 unique categories an item belongs to.
    \item \textbf{Prices}: Prices of items, normalized to follow a $\mathcal{N}(0,1)$ distribution.
\end{itemize}

\textbf{Zvuk} tracks user song-listening activity over the same five-month period. It includes 244,673,551 events across 12,598,314 listening sessions.  These sessions, initiated by 382,790 unique users, encompass 1,506,950 individual tracks. This dataset is specifically tailored to music and excludes other forms of auditory content, such as podcasts or audiobooks.

\textbf{Content}:
\begin{itemize}
\item \textbf{User IDs}: Individual identifiers, with a count of 382,790 users.
\item \textbf{Session IDs}: IDs for users' listening sessions, totaling \\ 12,598,314.
\item \textbf{Datetimes}: Timestamps spanning from \\ 01.15.2023 to 14.05.2023.
\item \textbf{Track IDs}: Identifiers for the music tracks, encompassing 1,506,950 unique tracks.
\item \textbf{Play Durations}: Scaled durations of tracks played, considering tracks where at least 30\% of the song's duration was completed.
\end{itemize}

\subsection{Preprocessing}
\label{appendix:a1}
We focus on collaborative filtering, converting datasets into implicit feedback through threshold binarization.
For datasets with ratings from 0 to 5, we use a threshold of $3.5$ for positive feedback. For datasets with weights from 0 to 1, such as percentage-based data, we use a threshold of $0.3$. Other datasets use thresholds based on the drop ratio.

For datasets like MegaMarket, which track user behavior via events, preprocessing is required to handle repeated user-item pairs. We preprocess the data in the following manner:
\begin{enumerate}
    \item Assign weights using the formula:
    \begin{equation}
        \frac{\sum\text{all\_interactions}}{\sum\text{type\_interactions}}
    \end{equation}
    where \textit{type\_interactions} denotes the count of events of a specific type. Thus, rarer events receive higher weights.
    \item Aggregate events by pairing users and items, retaining only the most frequent event type. We establish weight boundaries to ensure that events of a lesser significance never outnumber those of greater importance.
    \item Treat the newly assigned weights as ratings and apply threshold binarization accordingly.
\end{enumerate}

This preprocessing approach ensures consistent dataset handling in line with the principles of collaborative filtering.

\begin{figure*}[ht!]
\subfloat[Comparison of stability of aggregations with respect to a new similar method.\label{fig:add_similar_method}]
  {\includegraphics[width=.75\columnwidth]{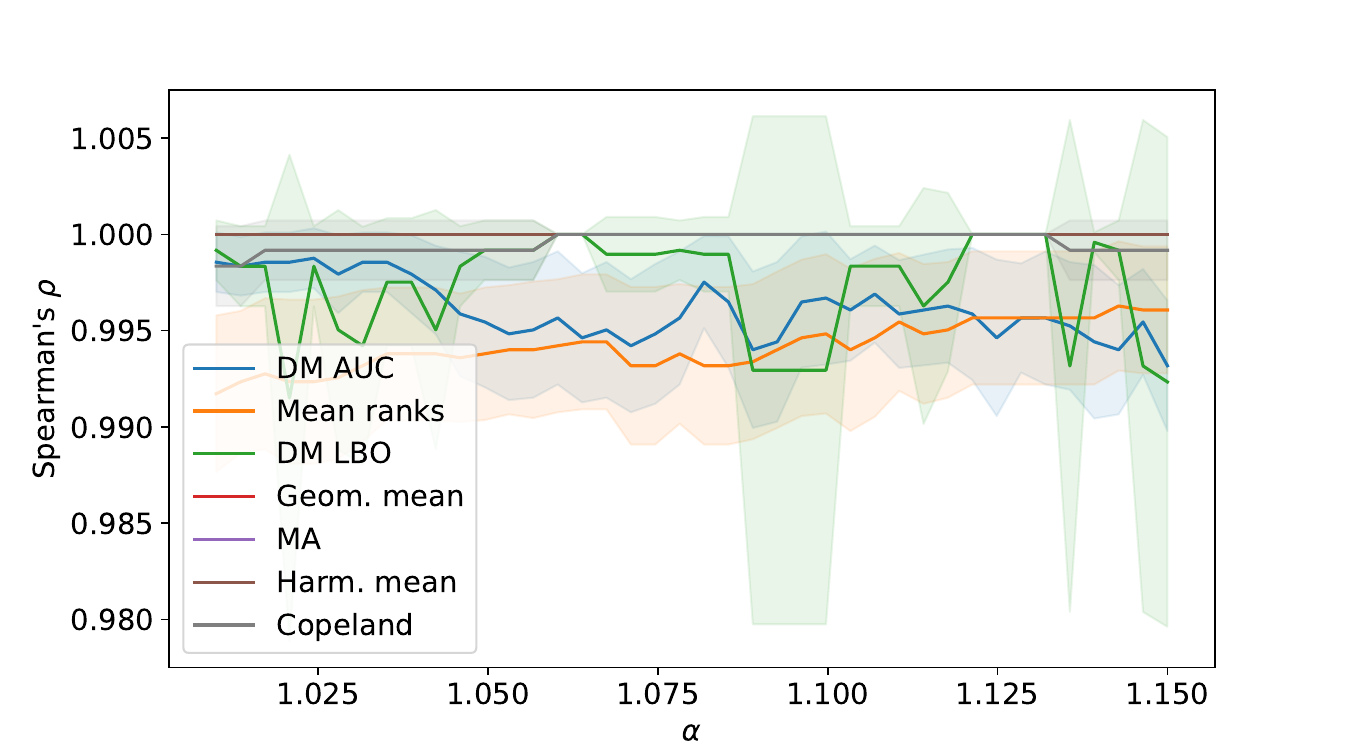}}\hfill
\subfloat[Comparison of stability of aggregations with respect to the new best method.\label{fig:add_method}]
  {\includegraphics[width=.59\columnwidth]{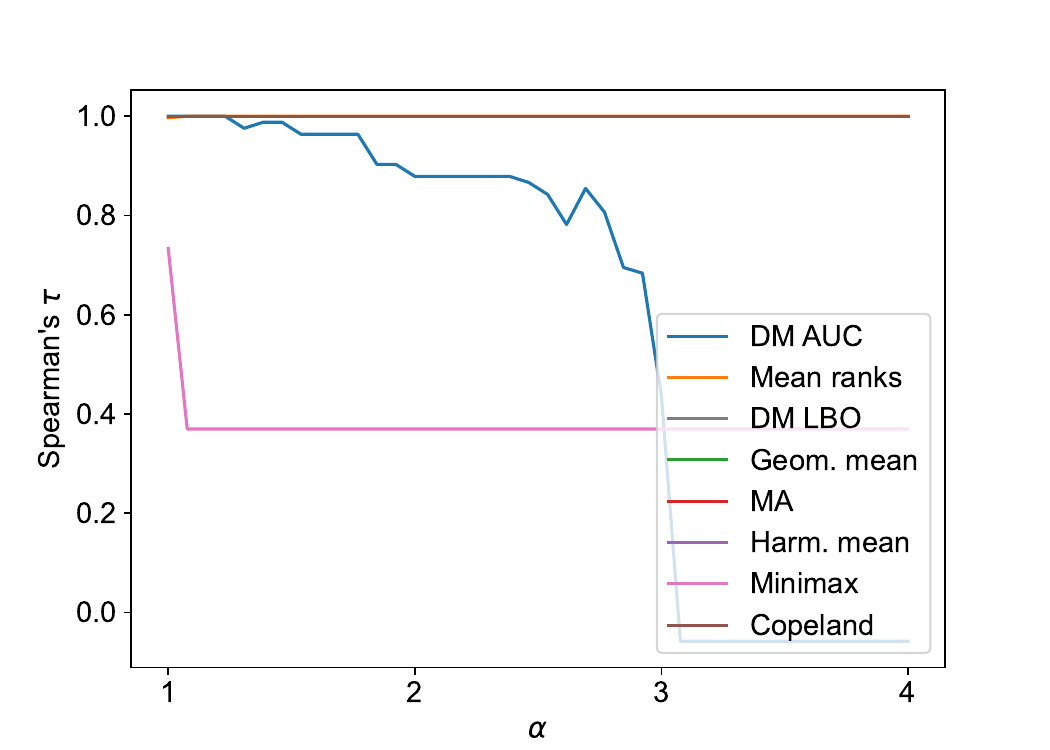}}\hfill
\subfloat[Comparison of DM aggregation approaches via Spearman correlation across different hyperparameters.\label{fig:hp_correlation_results}]
  {\includegraphics[width=.6\columnwidth]{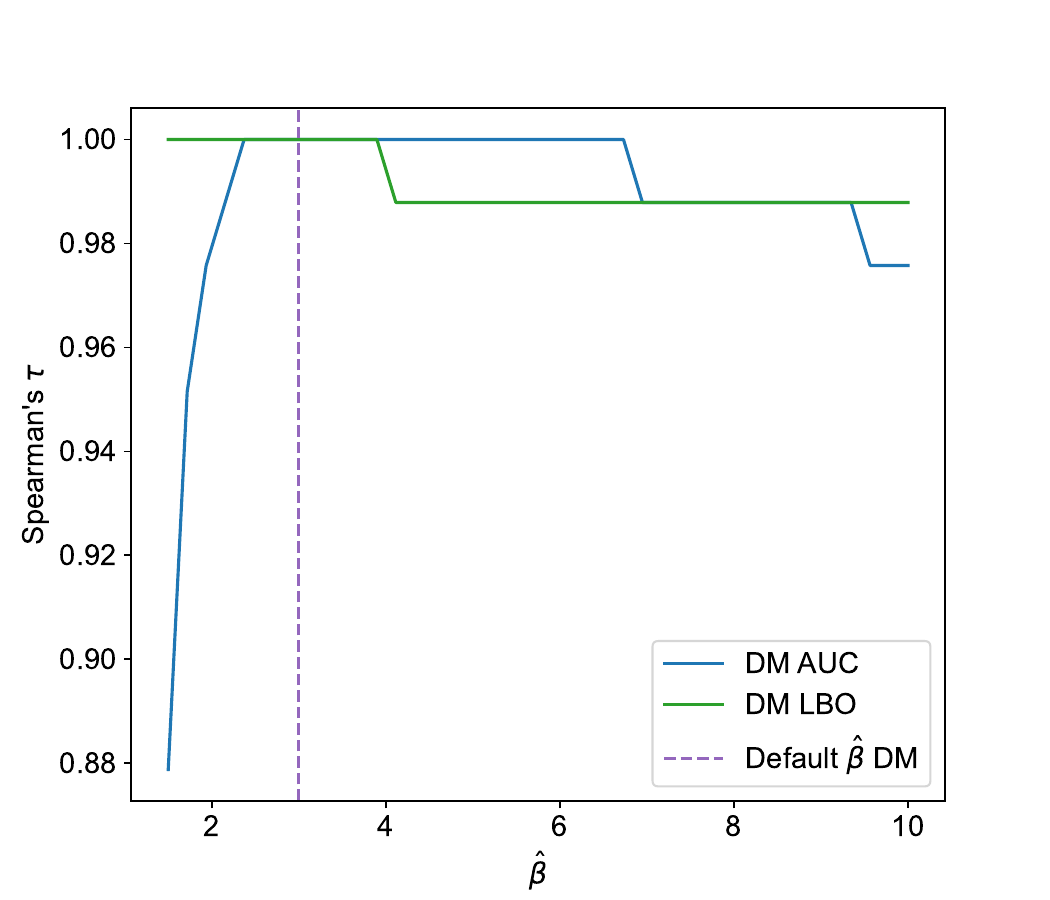}}
\caption{Reliability evaluation for various aggregation methods.}
\end{figure*}

\section{Additional Experiments}
\label{appendix:c}

\subsection{Reliability Tests: Incorporation or Exclusion of Marginally Different Methods}
\label{sec:slightly_inferior}
We examine the impact of introducing a model similar to existing ones, using a parameter $\alpha$ to adjust the metrics of a chosen model. We set  $|1 - \alpha| \leq 0.15$. For example, the case when $\alpha = 1.$ means adding the new model equal to the selected one.


Adjusting $\alpha$ makes the new model perform slightly better or worse than the chosen model. Results, shown in Figure~\ref{fig:add_similar_method}, indicate most aggregation methods are stable, except for \textbf{Mean ranks} and \textbf{Copeland}. The \textbf{Minimax} method is excluded due to poor Spearman correlation.


\subsection{Reliability tests: Adding a New Best Method}
\label{sec:cheat_method}
We analyze the impact of introducing a new best model. We identify the best metric values across all datasets and utilize an arbitrary value $\alpha$ (following similar conditions as in Section \ref{sec:slightly_inferior}). We restrict the parameter as follows: $\alpha \in [1, 4]$. For instance, when $\alpha = 1$, it signifies the addition of a new model with metrics equal to the current best metrics.

Figure~\ref{fig:add_method} shows the \textbf{Minimax} method is unstable for any $\alpha$, and \textbf{DM AUC} becomes unstable as $\alpha \rightarrow 4$. This behavior can be attributed to the direct dependence of the best metric values on the parameter $\alpha$. Conversely, all other aggregation methods remain entirely stable in this scenario.

\subsection{Sensitivity to Hyperparameters of Aggregations}
\label{sec:beta_sensitivity}


The results for varying the hyperparameters case are presented in Figure~\ref{fig:hp_correlation_results}. The vertical dotted line mean the case when $\hat{\beta} = 3$. We see that the \emph{DM AUC} aggregating method is more stable than the \emph{DM LBO} method. Instability of the \emph{DM AUC} can be interpreted by the proximity of the curves of the methods (in Figure~\ref{fig:dolan_more_curve}, for example, \emph{LightFM AUC} and \emph{MultiVAE AUC} curves look visually similar to each other). If the more methods have similar perfomance curves, then the more the Spearman correlation value decreases (in Figure~\ref{fig:dolan_more_curve}, for example, \emph{LightFM AUC}, \emph{MultiVAE AUC} and \textbf{ItemKNN AUC} curves for $\hat{\beta} \rightarrow 1. + 0$).

\section{Choosing the Optimal Subset of Datasets}
\label{appendix:optimal_design}

In the KMeans clustering process, we first standardize the data to ensure uniformity. Next, we apply Principal Component Analysis (PCA) to reduce the number of dimensions while preserving as much variance as possible, addressing the problem of correlated features. We then use the Isolation Forest method to detect and remove outliers, decreasing the dataset size from 30 to 25 observations. With the data now prepared, we move on to the clustering stage, employing the K-means algorithm. During clustering, we evaluate the Silhouette Coefficient and Davies-Bouldin Scores to determine the optimal number of clusters, which we found to be 6. Finally, we select datasets that are closest to the cluster centres for further analysis.

The next two approaches assume that resulting metrics can be predicted with the linear regression method:
\[
\mathbf{y} = \mathbf{x}\Theta + \varepsilon, \quad \varepsilon \sim \mathcal{N}(0, \sigma^2),
\]
where $\mathbf{y}$ is metric on datasets, $\Theta$ means model parameters, $\mathbf{x}$ means datasets feature and $\varepsilon$ is a noise value.

The Maximum Likelihood Estimation weights can be shown as
\[ \Theta \sim \mathcal{N}\left((\mathbf{x}^T\mathbf{x})^{-1}\mathbf{x}^\mathbf{x}\Theta, (\frac{1}{\sigma^2}\mathbf{x}^T\mathbf{x})^{-1}\right)
\]

The D-optimality approach we formulate as maximization of $\mathbf{x}^T\mathbf{x}$ 
determinant. As a result, we minimize the estimation variance.

We formulate the A-optimality as a result of the minimization of mean model loss. 
In this case, we assume that the prior distribution is standard normal distribution $p(x) = \mathcal{N}(0, I)$
\[
Q(\mathcal{D}) = \int (y(x) - \hat{y}(x))^2p(x)dx = \frac{1}{3}\text{tr}((\mathbf{x}^T\mathbf{x})^{-1})
\]

These problems are discrete optimization, and the naive solution is a Greedy algorithm. After selecting the starting datasets' subset, we perform a complete search across each dataset, choosing datasets with high values of the target function.

\end{document}